\def\a{\alpha}
\def\b{\beta}

\def\l{\lambda}

\def\ue3{\left| U_{e3} \right|}

\def\be{\begin{equation}}
\def\ee{\end{equation}}

\newcommand{\symmatrix}[6]{ \left( \!
                             \begin{array}{ccc}
                              #1 & #2 & #3 \\[-0.885pt]
                                 & #4 & #5 \\[-0.885pt]
                                 &    & #6 \\[-1.77pt]
                             \end{array} \!
                            \right) }

\documentstyle [12pt,epsfig]{article}
\begin{document}
\baselineskip=24 pt
\setcounter{page}{1}
\thispagestyle{empty}
\topskip 0.5  cm
\begin{flushright}
\begin{tabular}{c c}
& {\normalsize OCHA-PP-248}\\
& {\normalsize NIIG-DP-05-1}\\
&  July  2005
\end{tabular}
\end{flushright}

\thispagestyle{empty}
\vspace{0.5in}
\begin{center}
{\LARGE	\bf  Hybrid Textures  of   Neutrinos \\}
\vspace{1.5 cm}
{\bf Satoru Kaneko$^1$,  Hideyuki Sawanaka$^2$ and Morimitsu Tanimoto$^3$\\}
\vspace{0.2in}
{\sl $^1$ Physics Department, Ochanomizu University, Tokyo 112-8610, Japan\\}
{\sl $^2$ Graduate School of Science and Technology, Niigata University,
 Niigata 950-2181, Japan\\}
{\sl $^3$ Physics Department, Niigata University, Niigata 950-2181, Japan\\}
\vspace{1.5in}
\end{center}

\begin{abstract}
We present  numerical and comprehensive analyses of the sixty  hybrid textures
of neutrinos, which have an  equality of matrix elements and  one zero. 
These textures are possibly derived  from the discrete symmetry.
Only six textures among sixty ones  are excluded by the present experimental
data. Since there are many textures  which give similar predictions,
the textures are classified  based on  the numerical results.
The neutrinoless double beta decay is also examined in these textures.
Our results suggest  that there remain still rich structures of 
 the neutrino mass matrix in the phenomenological point of view.
\end{abstract}

\newpage
\baselineskip 24pt

\section{Introduction}

The neutrino mass-squared differences and the mixing angles
are now quite well determined by the neutrino oscillation experiments
\cite{solar,Kamland,atmospheric}.
The third mixing angle, represented by the matrix element $U_{e3}$
of the lepton mixing matrix $U$ (MNS matrix~\cite{mns}),
is  constrained to be small by the non-observation of neutrino oscillations
at the CHOOZ experiment~\cite{chooz}.

These results indicate the bi-large flavor mixing in the lepton sector. 
It is therefore important to investigate how the textures of lepton mass 
matrices can link up with the observables of the flavor mixings.
 Many authors studied the texture zeros
 \cite{Pre, FGM1, Xing, FGM2}, which may follow from the flavor  symmetry.
 On the other hand,  one finds  some  relations
among the non-zero mass matrix elements 
in the discrete symmetry of the flavor.
This fact suggests that the texture zero analyses is not enough to reveal
  some underlying flavor symmetry.

For example, in the $e, \mu, \tau$ basis,
 one finds the following symmetric mass matrices of neutrinos, where
there are  two entries with same values and one zero,
\begin{equation}
\left (\matrix{ a &  c & d\cr  & b & 0 \cr &  & b\cr}
\right ) \ , \qquad 
\left (\matrix{ 0 &  c & d\cr  & a & b \cr  &  & a\cr}
\right ) \ ,
\end{equation}
\noindent
both are presented  in the quaternion family symmetry, $Q_8$ \cite{Q8}
 and the latter is given   in the $S_3$ symmetry \cite{S3}.
The variant of these textures is also discussed in ref. \cite{Koide}.
We call this type texture as the  ``Hybrid'' texture,
which has an   equality of matrix elements and  one zero. 

The analytical study of various structures of the neutrino mass matrix was
presented systematically by Frigerio and Smirnov \cite{FS}, who also discussed
the case of equalities of matrix elements.
The textures for the Dirac neutrinos also discussed in ref. \cite{Hage}.
However, numerical and comprehensive analyses have not been given.  

In this paper, we present numerical  analyses of 
the  sixty hybrid textures,  which have 
  two equal mass matrix elements with one zero.
Our analyses include textures in the previous studies in refs. \cite{Q8,S3}.
Our results  are consistent with their results.

 The hybrid textures are discussed  in section 2,
 and numerical results are presented in section 3.
Section 4 is devoted to the summary.


 \section{Hybrid Textures} 

Let us construct the neutrino mass matrix in terms of neutrino mass 
eigenvalues $m_1,m_2,m_3$, mixing angles  and CP violating phases.
Neutrino mass matrix  $M_\nu$ 
     in the basis where the charged lepton mass matrix  is diagonal 
 (flavor basis)  is  given as follows:
\begin{equation}
 M_\nu= P \ U^* \ M_{\rm diagonal}  \ U^\dagger \ P \ ,
\end{equation}
\noindent where  $M_{\rm diagonal}$ and $P$ are  the diagonal mass matrix and
  the diagonal phase matrix, respectively:
\begin{equation}
M_{\rm diagonal}=\left (\matrix{ \l_1&0&0\cr0& \l_2&0 \cr
  0 & 0 & \l_3\cr}\right ) \ , \qquad\qquad
  P = \left (\matrix{ e^{i \phi_e}& 0 & 0 \cr  0 & e^{i \phi_\mu}&   0 \cr
  0 & 0 & e^{i\phi_\tau} \cr}\right )\ ,
\end{equation}
\noindent
where $\l_i\ (i=1,2,3)$ are the complex mass eigenvalues including 
Majorana phases, so three neutrino masses $m_i$ are given as 
the absolute values of $\lambda_i$. 
On the other hand, $\phi_i\ (i=e,\mu,\tau)$ are unphysical phases 
depending on the phase convention. The MNS mixing matrix $U$  is 
parametrized  as 
\begin{eqnarray}
U =\left (\matrix{c_{13}c_{12}& c_{13} s_{12} & s_{13}e^{-i \delta}\cr 
  -c_{23}s_{12}-s_{23}s_{13}c_{12}e^{i \delta}
 & c_{23}c_{12}-s_{23}s_{13}s_{12}e^{i \delta} &   s_{23}c_{13} \cr
  s_{23}s_{12}-c_{23}s_{13}c_{12}e^{i \delta} 
& -s_{23}c_{12}-c_{23}s_{13}s_{12}e^{i \delta} & c_{23}c_{13} \cr}\right )\ ,
\label{parametrization}
\end{eqnarray}
\noindent where $c_{ij}$ and  $s_{ij}$ denote
 $\cos \theta_{ij}$ and $\sin \theta_{ij}$,  respectively.
Then, the neutrino mass matrix elements  ${(M _\nu)}_{\alpha\beta}$
are given in the flavor basis as
\begin{eqnarray}
 (M _\nu)_{\alpha\beta} =
 e^{i(\phi_\a+\phi_\b)}\sum _f^3 U _{\alpha f}^* U_{\beta f}^* \lambda _f = 
e^{i(\phi_\a+\phi_\b)}\ (U_{\alpha 1}^* U_{\beta 1}^* \lambda_1+U_{\alpha 2}^* U_{\beta 2}^*\lambda _2 + U_{\alpha 3}^* U_{\beta 3}^* \lambda _3)\ .
\end{eqnarray}
By use of these mass matrix elements, we can analyze the sixty  cases
in terms of mass eigenvalues and mixings.
These cases are combinations of the one zero textures  and 
the equal  elements:
 (i) fifteen  cases of the equal  elements (Type $\rm A\sim O$)
$(M _\nu)_{\a\b} = (M_\nu) _{\gamma\delta}$,
 (ii) six  cases of one zero (Type $\rm I\sim VI$)
$(M _\nu)_{\alpha\beta} = 0$.
These combinations give ninety textures, however, among them, thirty textures
have two zeros, which have been studied in details 
\cite{Pre,FGM1, Xing, FGM2}.
Therefore, we study numerically in the sixty textures, 
which are summarized in Table 1.

\section{Numerical Results}
Since the textures have the conditions  $(M_\nu)_{\a\b}=0$ and
 $(M_\nu)_{ij}=(M_\nu)_{k\ell}$,
 we can get the ratios of mass eigenvalues by
solving two equations as follows \cite{Xing}:
\begin{eqnarray} 
\frac{\lambda_1}{\lambda_2} & = & 
\frac{(U_{i 3}^*U_{j 3}^*- Q\ U_{k3}^*U_{\ell 3}^*)  U_{\a 2}^* U_{\b 2}^* -
 (U_{i 2}^*U_{j2}^*- Q\ U_{k2}^*U_{\ell 2}^*)  U_{\a 3}^* U_{\b 3}^*}
{(U_{i 1}^* U_{j 1}^*- Q\ U_{k 1}^* U_{\ell 1}^*)  U_{\a 3}^* U_{\b 3}^* -
 (U_{i 3}^* U_{j3}^*- Q\ U_{k3}^* U_{\ell 3}^*)  U_{\a 1}^* U_{\b 1}^*}\ ,
\nonumber \\ \nonumber \\
\frac{\lambda_3}{\lambda_2} & = & 
\frac{(U_{i 1}^*U_{j 1}^*- Q\ U_{k1}^*U_{\ell 1}^*)  U_{\a 2}^* U_{\b 2}^* -
 (U_{i 2}^*U_{j2}^*- Q\ U_{k2}^*U_{\ell 2}^*)  U_{\a 1}^* U_{\b 1}^*}
{(U_{i 1}^* U_{j 1}^*- Q\ U_{k 1}^* U_{\ell 1}^*)  U_{\a 3}^* U_{\b 3}^* -
 (U_{i 3}^* U_{j3}^*- Q\ U_{k3}^* U_{\ell 3}^*)  U_{\a 1}^* U_{\b 1}^*}\ ,
\label{ratio}
\end{eqnarray}
 \noindent 
 where $ Q\equiv e^{i\varphi}= e^{i (\phi_k+\phi_\ell-\phi_i-\phi_j)}$.
Taking absolute values of these ratios, we get
the neutrino mass ratios, $m_1/m_2$ and $m_3/m_2$.
Therefore mass ratios are given in terms of  $\theta_{12}$,  $\theta_{23}$,
 $\theta_{13}$, $\delta$ and the unknown phase $\varphi$.
Absolute values of neutrino masses are fixed by putting the 
experimental data $\Delta m_{\rm atm}^2$ and $\Delta m_{\rm sun}^2$.

We present  the scattered plots of the allowed region in the
 $\sin ^2 2\theta_{23}-|U_{e3}|$ plane and the $m_2-m_3$ plane, 
in which the experimental data
are taken  in the $90\%$ CL limit \cite{Kamland,lisi}:
\begin{eqnarray}
 0.33 \le  \tan^2 \theta_{\rm sun} \le 0.49  \ , \ \
&&7.7\times 10^{-5} \le  \Delta m_{\rm sun}^2 \le
8.8\times 10^{-5}~\rm{eV}^2 \ , \nonumber \\
0.92 \le \sin^2 2\theta_{\rm atm}  \ , \ \
&& 1.5\times 10^{-3} \le  \Delta m_{\rm atm}^2 \le 3.4 \times 10^{-3}~
 \rm{eV}^2  \ .
\label{data}
\end{eqnarray} 
The bound obtained  by  the reactor neutrinos \cite{chooz} 
$\theta_{\rm reactor}\leq 12^\circ$ is also taken in our study.

 We classify  the textures based  on the  predicted mixings
  $\theta_{23}$ and $|U_{e3}|$.
 We cannot  distinguish the textures  by the mixing $\theta_{12}$
 at the present stage of the experimental data: 
\begin{itemize}
\item
The predicted mixings of the eighteen textures cover whole experimental 
allowed region:

 A-I, A-II, A-III, B-I, B-II, B-III, D-V, D-VI, E-II, F-I, I-I, I-II, 
   L-I, L-II, L-IV,  O-I, O-II, O-IV

\item
The allowed  points of the twenty-two textures  are very few:

   C-I, C-II, C-III, D-III,  E-V, E-VI, F-V, F-VI, G-III, G-V, 
G-VI, H-III, H-V, H-VI,  J-III, J-VI,  
K-III, K-VI, M-III, M-V,  N-III, N-V

\item
The  $\sin^2  2\theta _{23}$ has the lower bound $0.99$: 

 C-IV

\item
 The  $|U_{e3}|$ of the six textures has the lower bound, which increases 
 as $\sin^2 2\theta_{23}$ increases:

 A-VI, B-V,  G-II, H-I,  L-VI, O-V
 
The lower bound $|U_{e3}|\geq 0.03$ is   obtained 
 in  A-VI and  B-V, and 
 $|U_{e3}|\geq 0.04$  is  clearly predicted
in   G-II, H-I.
The bound $|U_{e3}|\geq 0.05$ is roughly obtained 
 although the allowed points are few  in    L-VI and  O-V.

\item
 The  lower bound of $|U_{e3}|$ decreases
 as $\sin^2 2\theta_{23}$ increases in the seven textures:

  D-IV,  I-V, I-VI, J-II, K-I, M-II,  N-I,

in which $|U_{e3}|=0$ is allowed  at $\sin^2 2\theta_{23}=1$
 except D-IV.  The texture  D-IV has the lower bound $|U_{e3}|\geq 0.001$.

\item
The six textures are  excluded by the experimental data:

 E-IV,  F-IV,  J-IV, K-IV, M-IV, N-IV
\end{itemize}

The mass patterns are  summarized in Table 2, in which
NH, IH and DL denote the normal mass hierarchy, inverted one and 
quasi-degenerate masses, respectively, and ALL means including
NH, IH and DL.

For the typical textures in each group,
the allowed regions are shown by scattered  plots in both
 $\sin ^2 2\theta_{23}-|U_{e3}|$ plane and the $m_2-m_3$ plane
 in  fig. 1 $\sim$ fig. 5, where 200,000 (1,000,000)
 parameter sets are produced randomly for A-I, G-II, I-V (C-III, C-IV) and 
 only experimental allowed sets are plotted.

The texture A-I is a typical one, which leads to the  normal hierarchy of 
the neutrino masses mainly, but the quasi-degenerate spectrum is also allowed,
as seen in fig. 1.
The predicted mixings  cover all  experimental allowed
 region on  the  $\sin ^2 2\theta_{12}-|U_{e3}|$ plane
 as well as on  the  $\sin ^2 2\theta_{23}-|U_{e3}|$ plane.
The texture C-III  is a typical one, which leads to the inverted  
hierarchy of the neutrino masses mainly
as seen in fig. 2, in which  1,000,000 parameter sets are put randomly 
 since the allowed points are very few.
The texture C-IV  gives a specific mass hierarchy
 and  mixing angle $\theta_{23}$ as seen in fig. 3, on the other hand,
  the predicted $\theta_{12}$ covers whole experimental allowed region.
The texture G-II is a typical one, which gives also 
a specific mass hierarchy
 and the  clear lower bound of  $|U_{e3}|$ as seen in fig. 4.
The texture I-V is a typical one, which leads to 
the inverted mass hierarchy of neutrino masses. The prediction excludes
the  specific region  on  the  $\sin ^2 2\theta_{23}-|U_{e3}|$ plane
  as seen in fig. 5.
We also show the relative magnitudes of the matrix elements
 in the unit of $\lambda\simeq 0.2$ for these typical textures  in Table 3.

It may be helpful  to see which future data might rule out
these textures.
If the inverted mass hierarchy is shown to be realized by Nature,
 the textures of C-IV and G-II are ruled out.
On the contrary, if mass spectrum is the normal hierarchy,
the textures of  C-III and I-V are ruled out.
Finding $\sin^2 2\theta_{23}<0.98$ and $|U_{e3}|<0.04$ rule out
the texture C-IV and G-II, respectively.

Our numerical results can be understandable 
in the following  analytical studies by using eq.(\ref{ratio}).
For the texture C-IV, the mass ratios $m_1/m_2$ and $m_3/m_2$ are given as
\begin{eqnarray} 
&&\frac{m_1}{m_2}=\left |\frac{\lambda_1}{\lambda_2}\right |=
\frac{s_{12}^2}{c_{12}^2} \ , \nonumber\\
&&\frac{m_3}{m_2}=\left |\frac{\lambda_3}{\lambda_2}\right |=
\frac{\cos 2\theta_{12}}{c_{12}^2} \left |\  1-
\frac{1}{2}s_{13}\tan 2\theta_{12}\sin 2\theta_{23}
\frac{(1+Q)e^{-i\delta}}{c_{23}^2 Q -s_{23}^2} \ \right |\ ,
\label{c4}
\end{eqnarray} 
\noindent where the first order of $s_{13}$ is remained and the higher order 
 $s_{13}^2$ are neglected.
It is found that  one obtains $m_1/m_2\simeq 2/3$ and $m_3/m_2\simeq 1/3$ 
 in the limit of $s_{13}=0$ with putting $\theta_{12}=\pi/6$ and  
$\theta_{23}=\pi/4$.  The value of $m_1/m_2\simeq 2/3$ indicates 
the normal mass hierarchy. However,  $m_3/m_2\simeq 1/3$ contradicts with 
the experimental mass ratios, and so it is concluded that
  $s_{13}=0$ is forbidden.
In order to see the $s_{13}$ dependence of the lower bound of 
$\sin^2 2\theta_{23}$, we take a parametrization of 
$\theta_{23}=\pi/4-\epsilon$. Putting it on  eq.(\ref{c4}), we get
\begin{eqnarray} 
\frac{m_3}{m_2}=
\frac{\cos 2\theta_{12}}{c_{12}^2} \left |\  1-
\frac{1}{2}s_{13}\tan 2\theta_{12}\
\frac{(1+Q)e^{-i\delta}}{\epsilon (1+Q)+\frac{1}{2}(Q-1)} \ \right |\ ,
\end{eqnarray} 
\noindent
in the first order of $\epsilon$.  The largest $m_3/m_2$ is obtained
in the case of $Q=1$ and $e^{-i \delta}=-1$.  With these phases,
$\epsilon$ is given in terms of $s_{13}$ and $m_3/m_2$ as
\begin{eqnarray} 
\epsilon=\frac{\tan 2\theta_{12}}{2}s_{13}
\left(\ \frac{m_3}{m_2}\frac{c_{12}^2}{\cos 2\theta_{12}} -1 \right )^{-1}\ .
\end{eqnarray} 
Taking the boundary of  data in eq.(\ref{data}),
 one can  get the upper  bound of  $\epsilon$ for the fixed  $s_{13}$.
 In the case of $s_{13}=0.2$, we obtain $\epsilon=0.04$, which leads
 to the lower bound $\sin^2 2\theta_{23}=0.99$. This result
is good agreement with the numerical one in fig. 3.

It is more easy to understand numerical results for the texture  G-II.
 The mass ratio $m_3/m_2$ is  given as
\begin{eqnarray} 
\frac{m_3}{m_2}=\left |\frac{\lambda_3}{\lambda_2}\right |=\left |
\frac{s_{12} s_{23}}{s_{13}(c_{12}c_{23}+Q s_{12})} \right | \ ,
\label{g2}
\end{eqnarray} 
\noindent which is larger than 1 due to $s_{13}\leq 0.2$.
Therefore, this texture corresponds to the normal hierarchy of the mass.
The lower bound of $s_{13}$ is given with $Q=1$ as
\begin{eqnarray} 
s_{13} \geq \frac{m_2}{m_3}\
\frac{s_{12} s_{23}}{s_{13}(c_{12}c_{23}+ s_{12})} \ .
\end{eqnarray} 
\noindent 
Taking the boundary of  data in eq.(\ref{data}), we obtain
the lower bound $s_{13}=0.04$, which is also  good agreement
 with the numerical one in fig. 4.

For the texture I-V, 
the normal mass hierarchy is not allowed as seen in fig. 5.
 The mass ratio  $m_3/m_2$ is  given as
\begin{eqnarray} 
\frac{m_3}{m_2}=
\left |\frac{c_{23}^2[\cos 2\theta_{12}e^{i\delta}
 - c_{12}s_{12}s_{13}(Q+2\tan\theta_{23})]}{s_{23}c_{23}s_{12}^2 Q e^{i\delta}+ c_{12}^2s_{23}^2 e^{i\delta}+s_{12}c_{12}s_{23}^2 s_{13} Q}\right |  \ .
\label{i5}
\end{eqnarray} 
In the limit of $s_{13}=0$, we have
\begin{eqnarray} 
\frac{m_3}{m_2}=
\left |\frac{c_{23}\cos 2\theta_{12}}{s_{23}\ (s_{12}^2 Q + 
c_{12}^2 \tan 2\theta_{23})} \right | \ ,
\end{eqnarray} 
\noindent 
while
\begin{eqnarray} 
\frac{m_1}{m_2}=
\left |\frac{c_{12}^2 Q + s_{12}^2 \tan \theta_{23}}{s_{12}^2 Q + 
c_{12}^2 \tan \theta_{23}} \right | \ ,
\end{eqnarray} 
\noindent at  $s_{13}=0$.
It is noticed that  the normal hierarchy mass is not allowed
   for an arbitrary phase $Q$, because  $m_3/m_2$ is not larger than 1.
Actually, in the case of $Q=-1$ and $\theta_{23}\simeq \pi/4$, 
neutrino masses
are degenerate, on the other hand, taking  $Q=-1$  
 we have $m_3/m_2<1 $, therefore,  the inverted mass hierarchy is 
realized. The situation is not changed even if we include
 the  small  $s_{13}$ effect.

\begin{figure}
\begin{center}
\epsfxsize=8.0 cm
\epsfbox{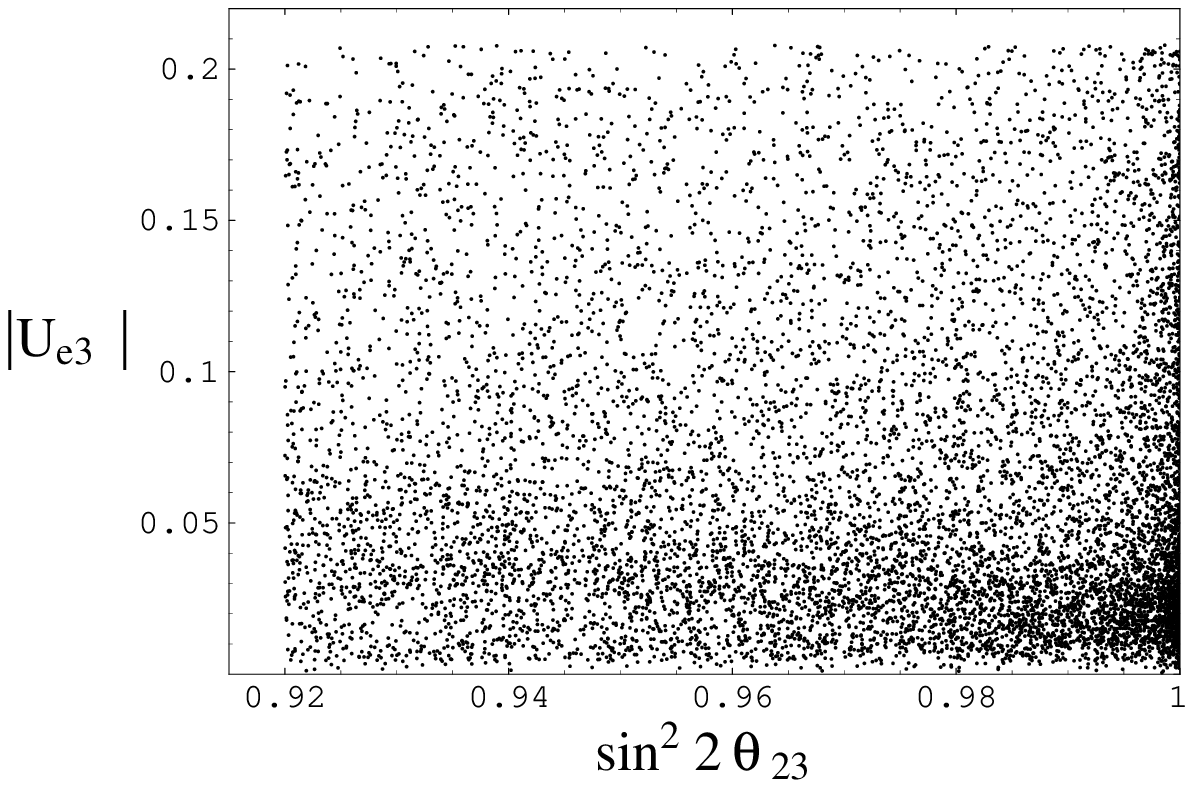}
\hskip 0.5 cm
\epsfxsize=7.0 cm
\epsfbox{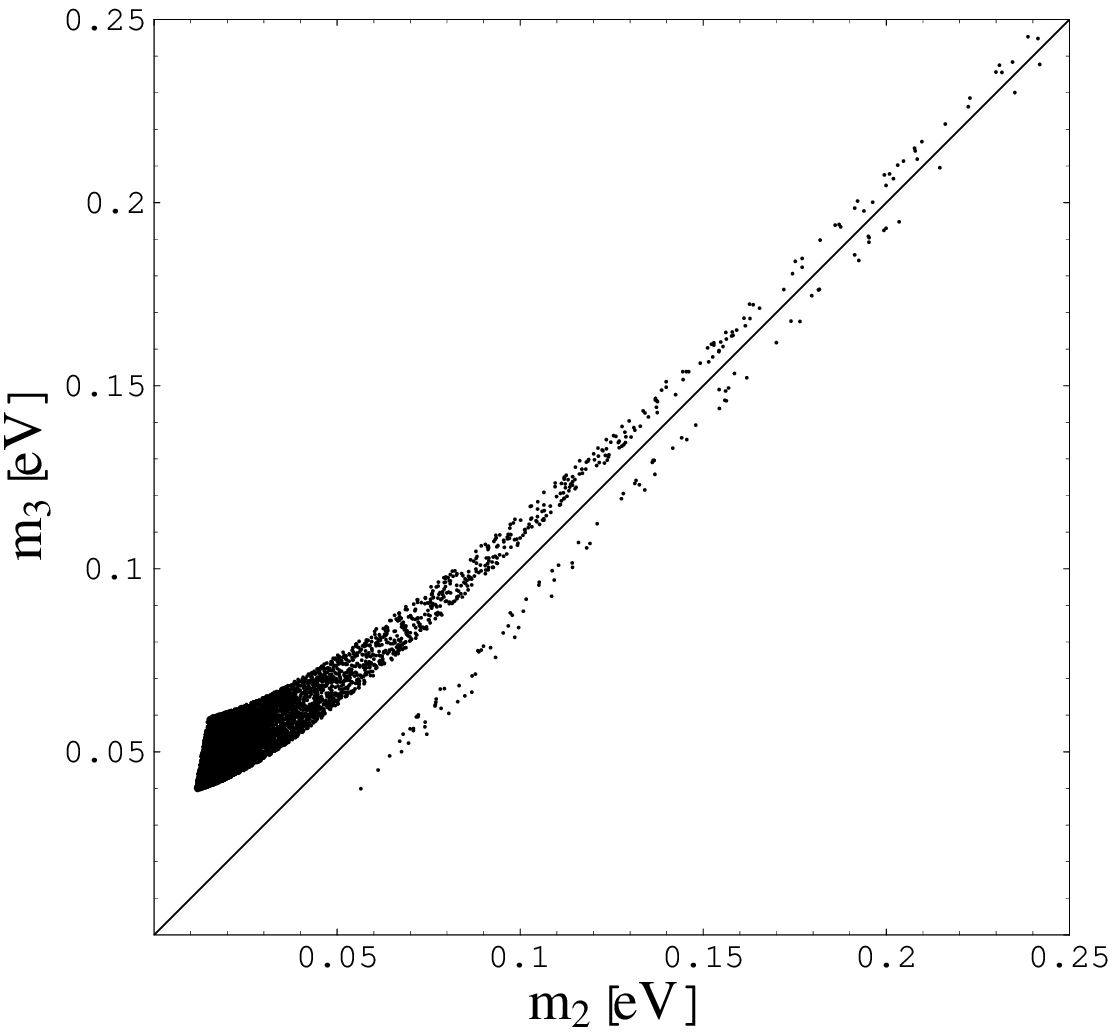}
\end{center}
\vspace*{.0cm}
\caption{Scattered plot of the allowed region in the texture A-I. }
\end{figure}%
\begin{figure}
\begin{center}
\epsfxsize=8.0 cm
\epsfbox{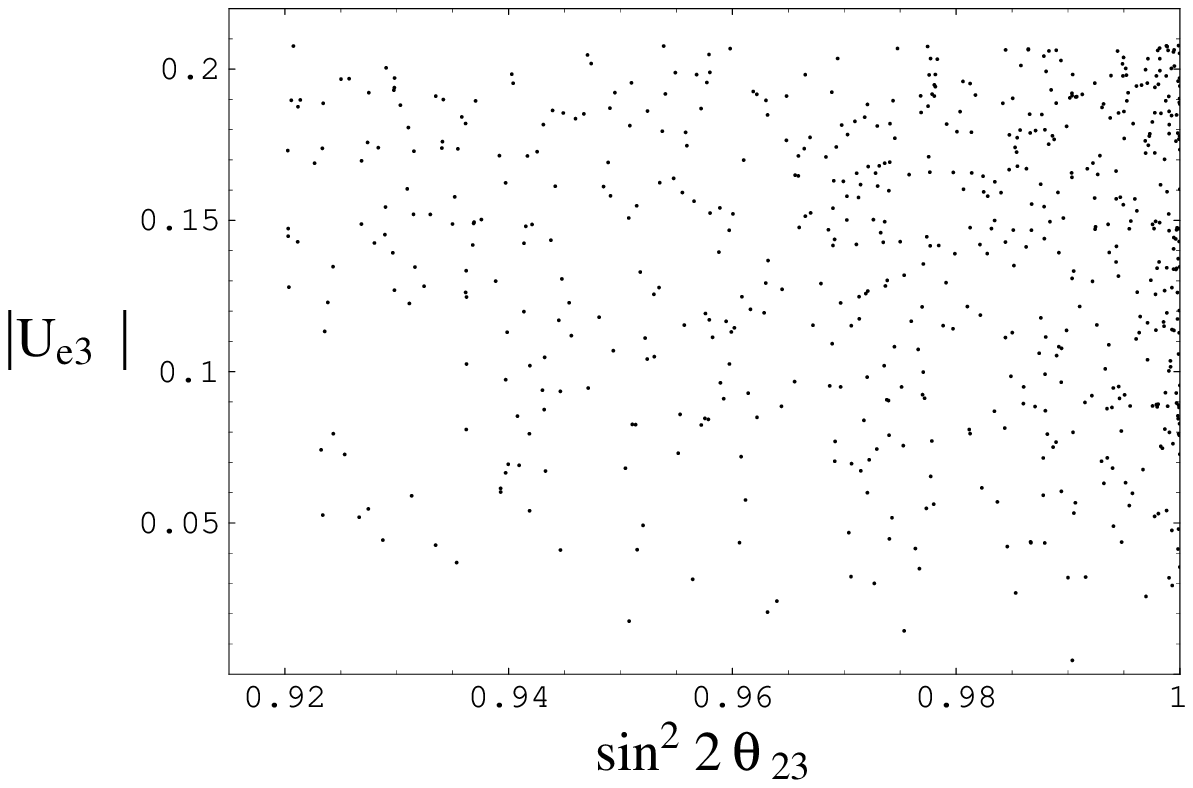}
\hskip 0.5 cm
\epsfxsize=7.0 cm
\epsfbox{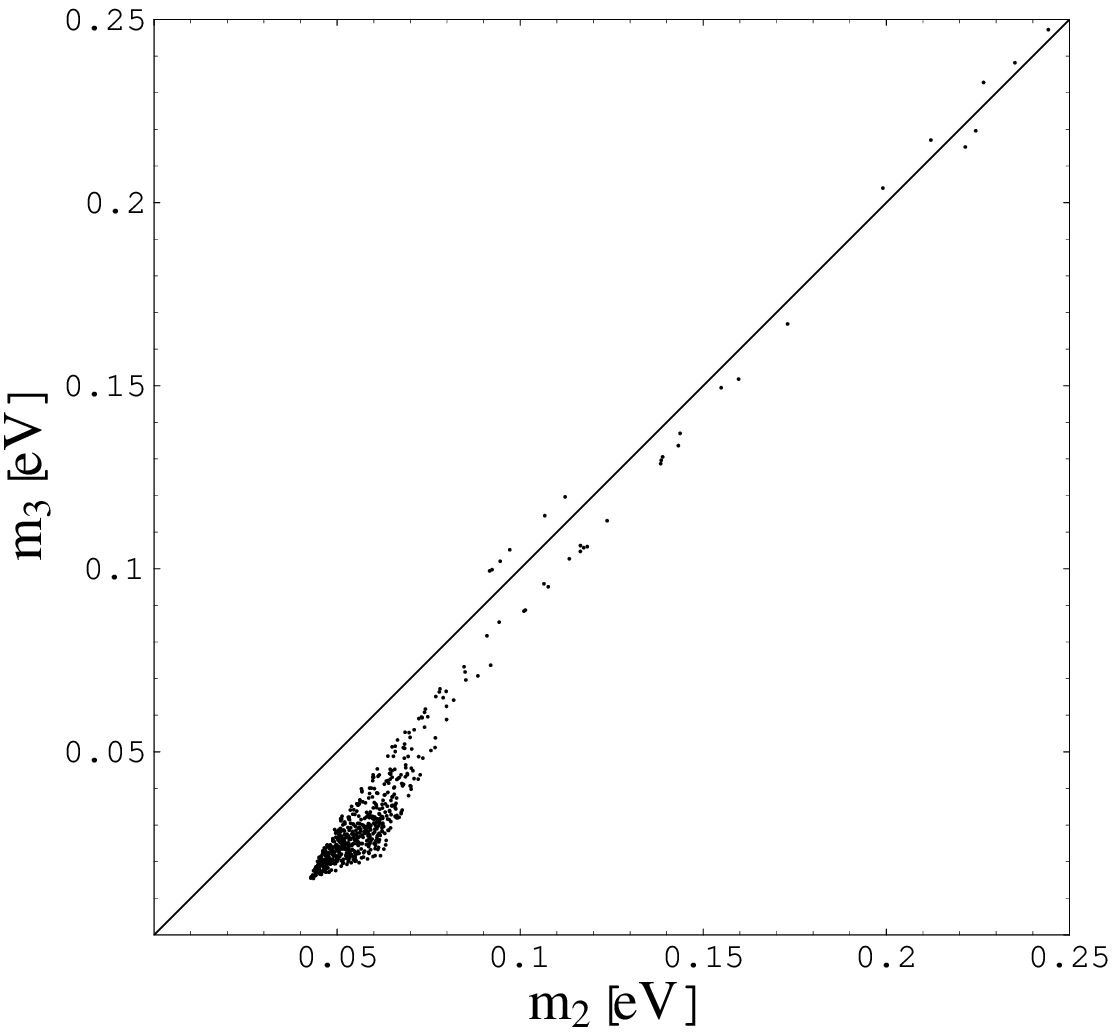}
\end{center}
\vspace*{.0cm}
\caption{Scattered plot of the allowed region in the texture C-III.}
\end{figure}%
\begin{figure}
\begin{center}
\epsfxsize=8.0 cm
\epsfbox{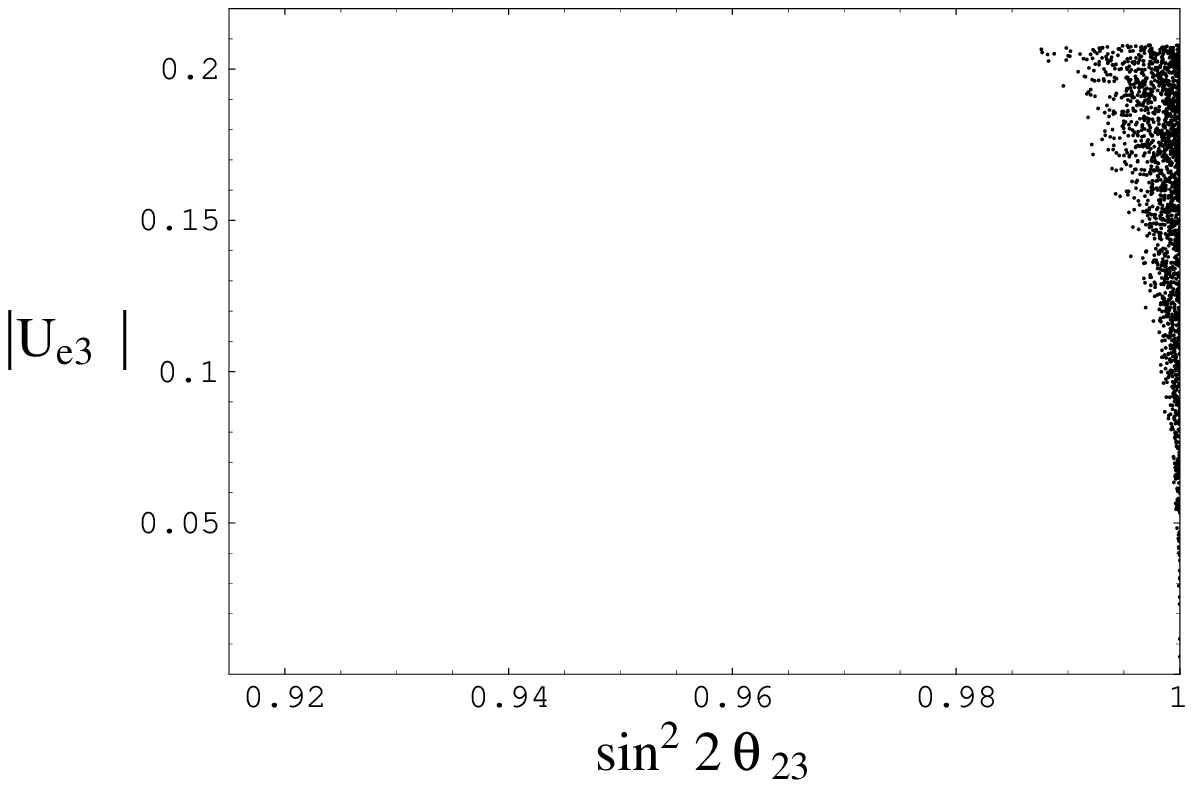}
\hskip 0.5 cm
\epsfxsize=7.0 cm
\epsfbox{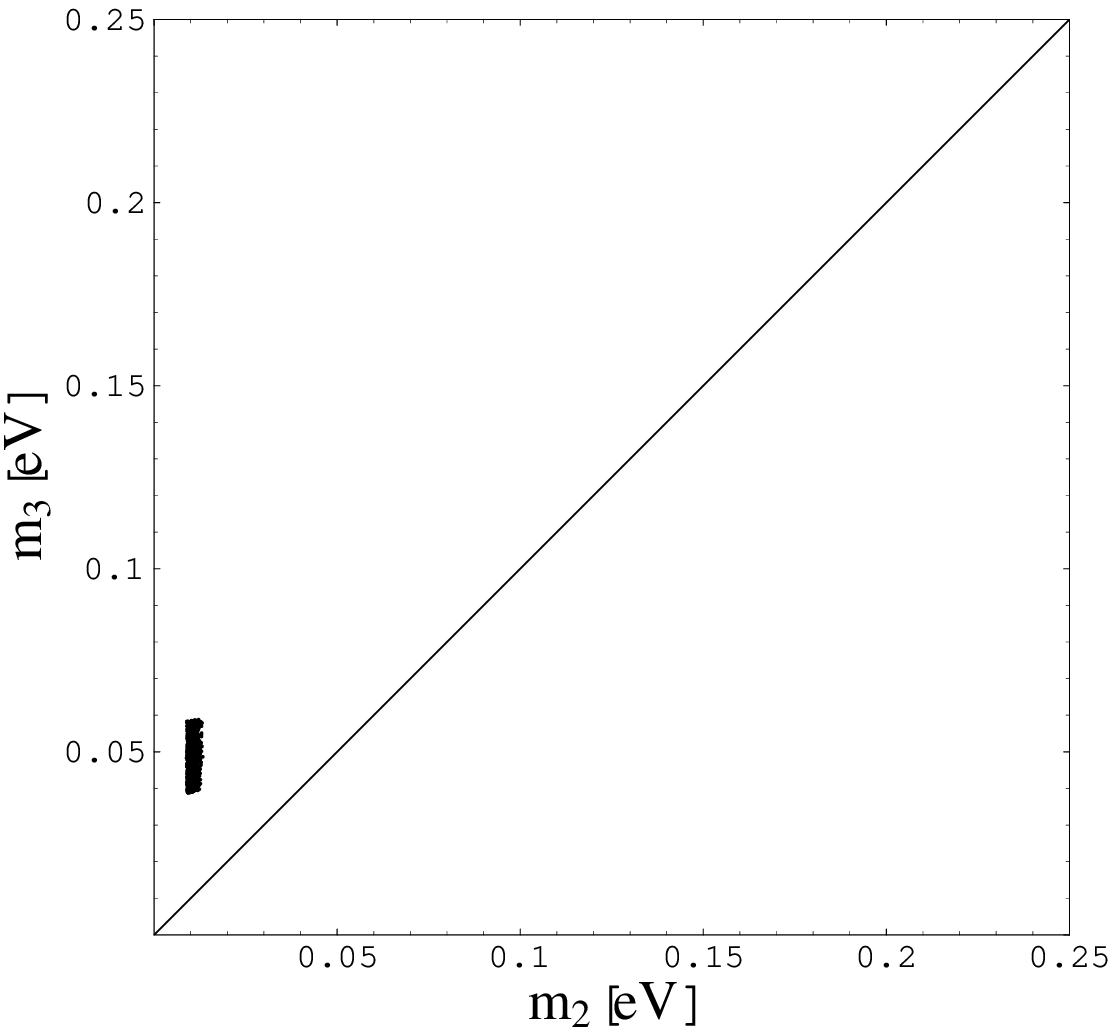}
\end{center}
\vspace*{.0cm}
\caption{Scattered plot of the allowed region in the texture C-IV.}
\end{figure}%
\begin{figure}
\begin{center}
\epsfxsize=8.0 cm
\epsfbox{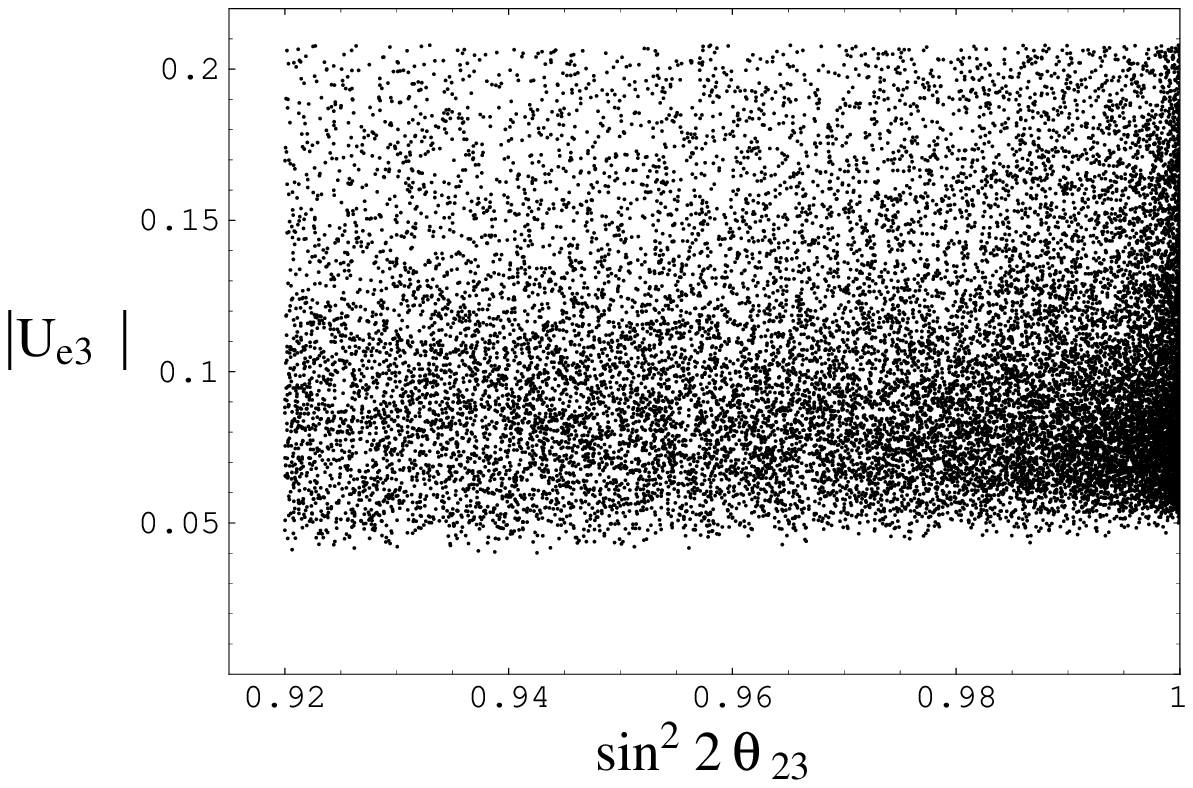}
\hskip 0.5 cm
\epsfxsize=7.0 cm
\epsfbox{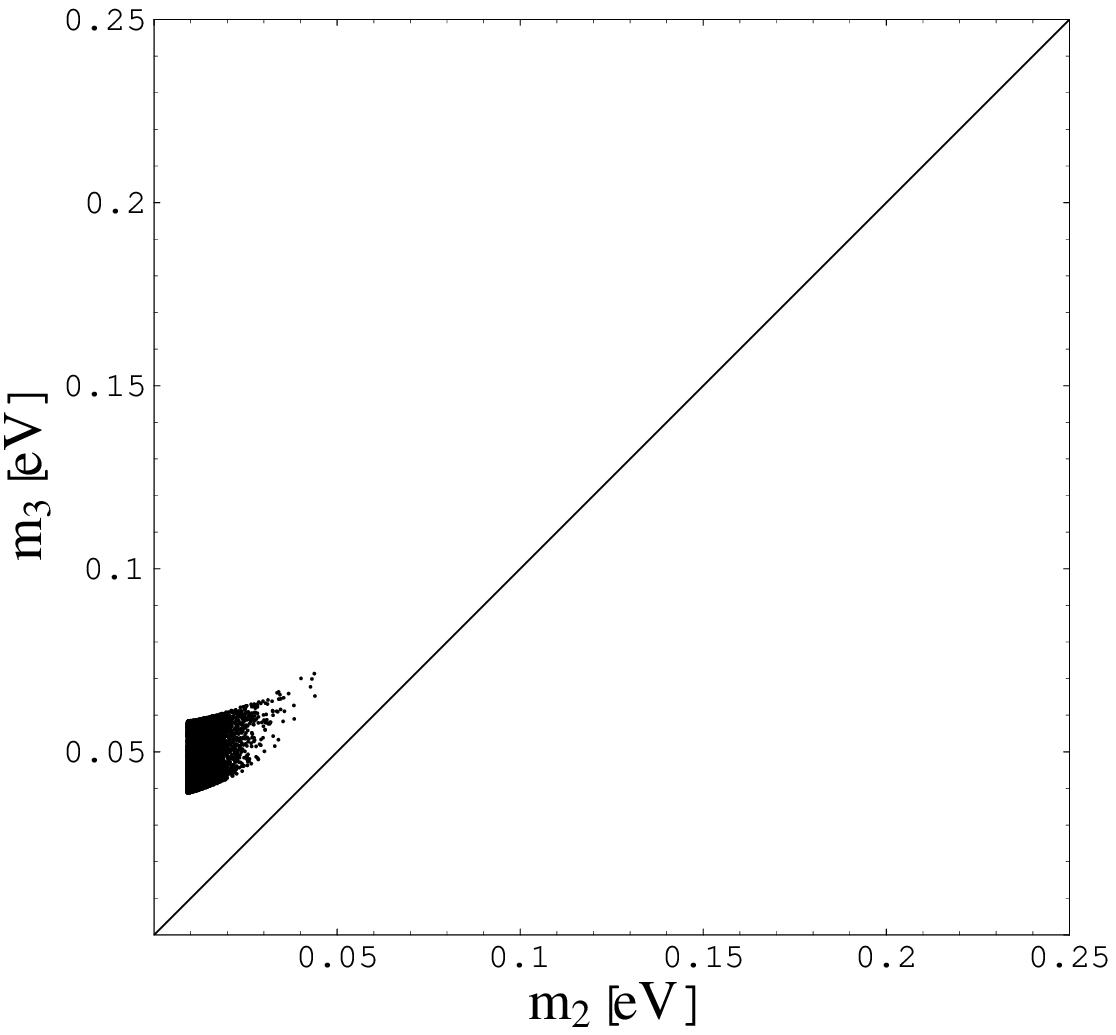}
\end{center}
\vspace*{.0cm}
\caption{Scattered plot of the allowed region in the texture G-II.}
\end{figure}%
\begin{figure}
\begin{center}
\epsfxsize=8.0 cm
\epsfbox{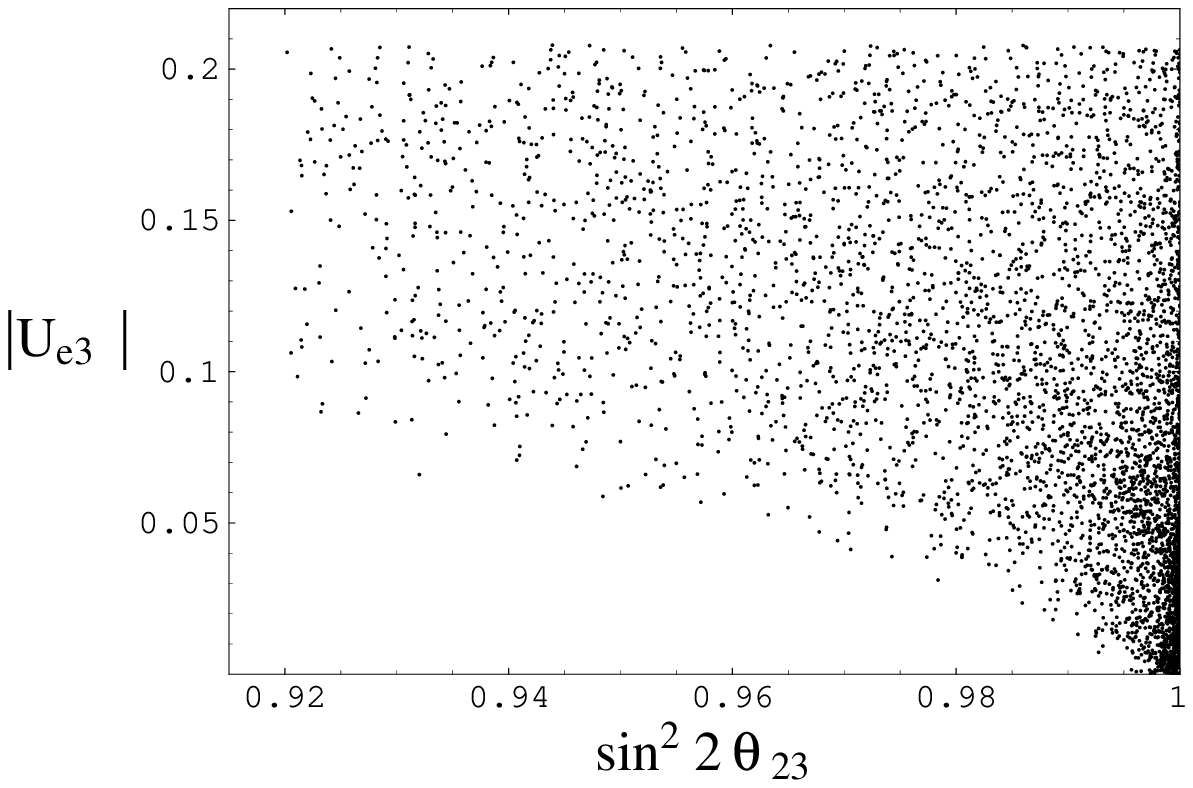}
\hskip 0.5 cm
\epsfxsize=7.0 cm
\epsfbox{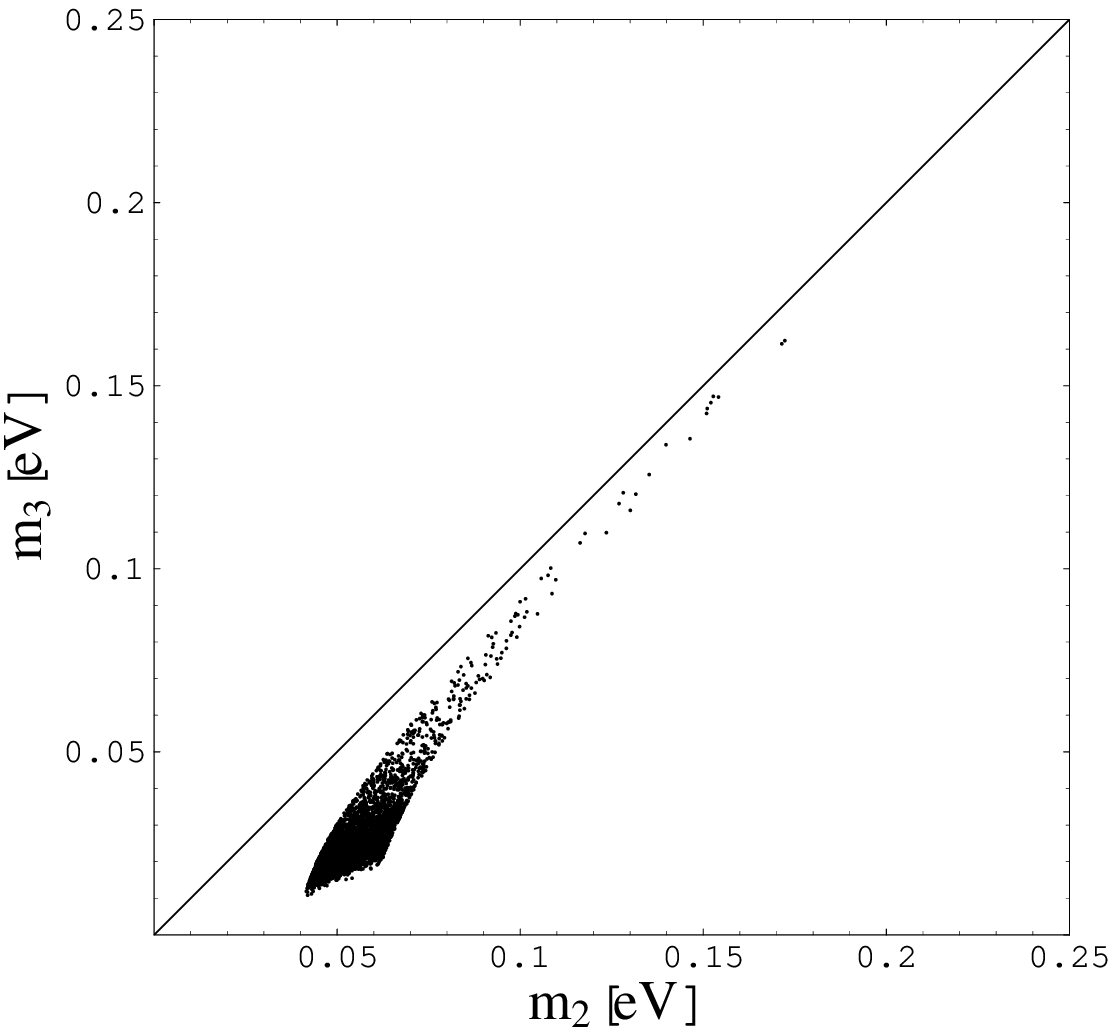}
\end{center}
\vspace*{.0cm}
\caption{Scattered plot of the allowed region in the texture I-V.}
\end{figure}%

Let us discuss the neutrinoless double beta decay rate, which 
 is controlled by the effective Majorana mass:
\begin{eqnarray}
 \langle m \rangle_{ee}=\left | \ m_1 c_{12}^2c_{13}^2 e^{i \rho} +
 m_2 s_{12}^2c_{13}^2 e^{i \sigma}+ m_3 s_{13}^2 e^{-2i \delta} \ \right | \ ,
\end{eqnarray}
\noindent where
$\rho=\arg{(\lambda_1/\lambda_3)}$ and $\sigma=\arg{(\lambda_2/\lambda_3)}$. 
This effective mass is just the absolute value of $(M_\nu)_{ee}$ component
 of the neutrino mass matrix. 

The predicted lower bounds of $\langle m \rangle_{ee}$   
are classified for  textures in Table 4.
It is remarked that the neutrinoless double beta decay is forbidden 
in the textures of type IV, because of $(M_\nu)_{ee}=0$.
The predictions of the typical textures D-V, E-II, L-II are presented
in figures 6, 7 and 8. 

 As shown in Table 4, many hybrid textures (thirty-eight ones) predict 
the lower bound $10\sim 30$ meV although there are differences of factor 
in  the lower bound predictions for each texture.
\begin{figure}
\begin{center}
\epsfxsize=7.  cm
\epsfbox{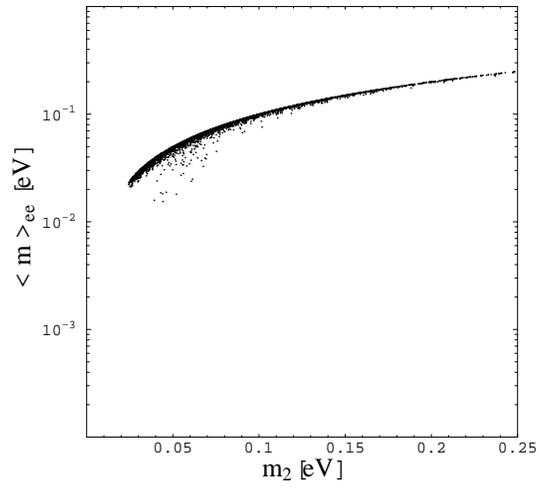}
\end{center}
\vspace*{-.1cm}
\caption{Neutrinoless $\b\b$ decays  in the texture D-V.}
\end{figure}%
\begin{figure}
\begin{center}
\epsfxsize=7. cm
\epsfbox{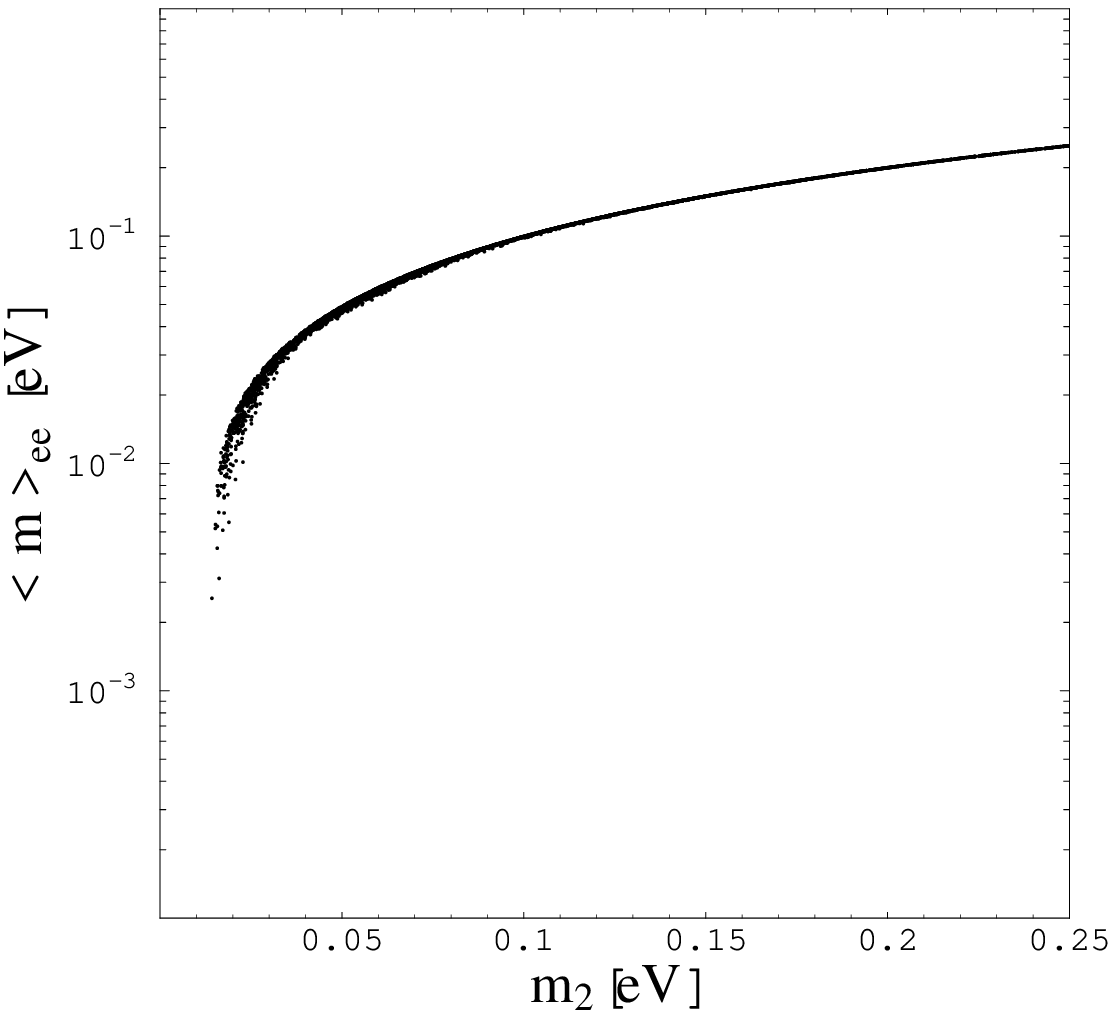}
\end{center}
\vspace*{-.1 cm}
\caption{Neutrinoless $\b\b$ decays  in the texture E-II.}
\end{figure}
\vskip -3 cm
\begin{figure}
\begin{center}
\epsfxsize=7.  cm
\epsfbox{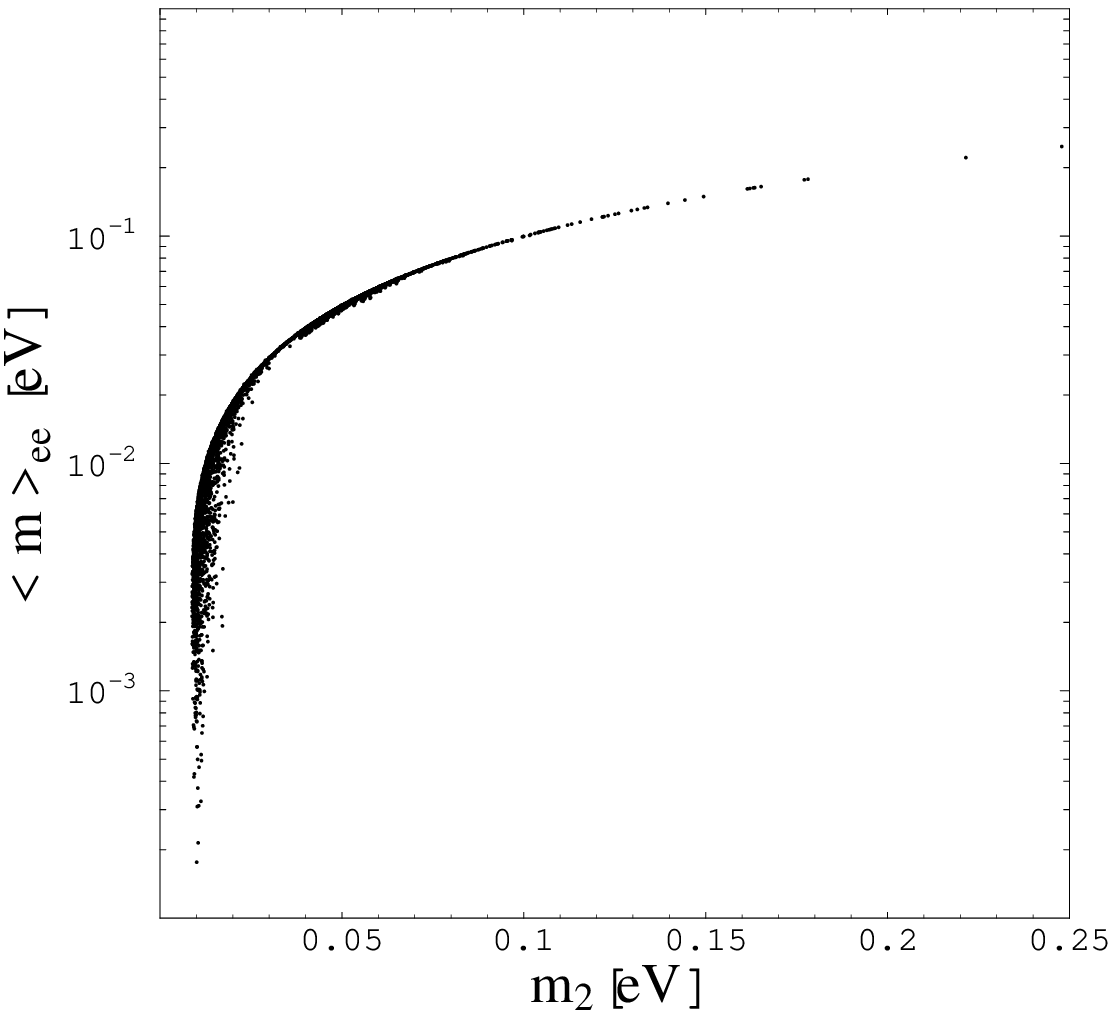}
\end{center}
\vspace*{.0cm}
\caption{Neutrinoless $\b\b$ decays  in the texture  L-II.}
\end{figure}
\newpage
It is remarked that there is 
the approximate  $\mu-\tau$ flavor interchange symmetry \cite{Sym23} 
in the textures.     Let us consider the following transformation:
\begin{eqnarray}
 S^T \ M_{\nu} \ S = \hat{M}_\nu \ , \qquad
S= \left ( \matrix{1 &0 & 0 \cr 0& 0 & 1 \cr 0& 1 & 0\cr
} \right ) \ ,
\end{eqnarray}
 which corresponds to $\theta_{23} \Leftrightarrow \frac{\pi}{2}-\theta_{23}$
 interchange.  Our results indicate ${M}_\nu$ and $\hat{M}_\nu$ are ``dual''. 
The two textures related by $S^TM_{\nu}S$
present almost same  results in the $\sin ^2 2\theta_{23}$ plot.
This symmetry is due to  $\theta_{23}\sim \pi/4$ experimentally. 
 The twenty-eight  sets of the   $\mu-\tau$ flavor interchange are given as
\begin{eqnarray}
 &&
 \textrm{A-I}   \Leftrightarrow \textrm{B-II},  \quad
 \textrm{A-II}  \Leftrightarrow \textrm{B-I},   \quad
 \textrm{A-VI}  \Leftrightarrow \textrm{B-V},   \quad 
 \textrm{A-III} \Leftrightarrow \textrm{B-III}, \nonumber\\
 &&
 \textrm{L-I}   \Leftrightarrow \textrm{O-II},  \quad
 \textrm{L-II}  \Leftrightarrow \textrm{O-I},   \quad
 \textrm{L-VI}  \Leftrightarrow \textrm{O-V},   \quad
 \textrm{L-IV}  \Leftrightarrow \textrm{O-IV},  \nonumber \\
 &&
 \textrm{E-II}  \Leftrightarrow \textrm{F-I},   \quad
 \textrm{E-V}   \Leftrightarrow \textrm{F-VI},  \quad
 \textrm{E-VI}  \Leftrightarrow \textrm{F-V},   \quad
 \textrm{E-IV}  \Leftrightarrow \textrm{F-IV},  \nonumber\\
 &&
 \textrm{G-II}  \Leftrightarrow \textrm{H-I},   \quad
 \textrm{G-V}   \Leftrightarrow \textrm{H-VI},  \quad
 \textrm{G-VI}  \Leftrightarrow \textrm{H-V},   \quad
 \textrm{G-III} \Leftrightarrow \textrm{H-III}, \nonumber\\
 &&
 \textrm{J-II}  \Leftrightarrow \textrm{N-I},   \quad
 \textrm{J-VI}  \Leftrightarrow \textrm{N-V},   \quad
 \textrm{J-III} \Leftrightarrow \textrm{N-III}, \quad
 \textrm{J-IV}  \Leftrightarrow \textrm{N-IV},  \nonumber\\
 &&
 \textrm{K-I}   \Leftrightarrow \textrm{M-II},  \quad
 \textrm{K-VI}  \Leftrightarrow \textrm{M-V},   \quad
 \textrm{K-III} \Leftrightarrow \textrm{M-III}, \quad
 \textrm{K-IV}  \Leftrightarrow \textrm{M-IV},  \nonumber\\
 &&
 \textrm{C-I}   \Leftrightarrow \textrm{C-II},  \quad
 \textrm{D-V}   \Leftrightarrow \textrm{D-VI},  \quad
 \textrm{I-I}   \Leftrightarrow \textrm{I-II},  \quad
 \textrm{I-V}   \Leftrightarrow \textrm{I-VI},
\end{eqnarray}
except for
\begin{eqnarray}
 \textrm{C-III}, \quad \textrm{C-IV}, \quad
 \textrm{D-III}, \quad \textrm{D-IV},
\end{eqnarray}
\noindent which are transformed to themselves.
In the fig. 9, we show the scattered plot on
the $\theta_{23}-|U_{e3}|$ plane  for the textures of  
A-I  and  B-II, in which   one can see the   interchange symmetry of 
$\theta_{23} \leftrightarrow \frac{\pi}{2}-\theta_{23}$.
\begin{figure}
\begin{center}
\epsfxsize=7.0 cm
\epsfbox{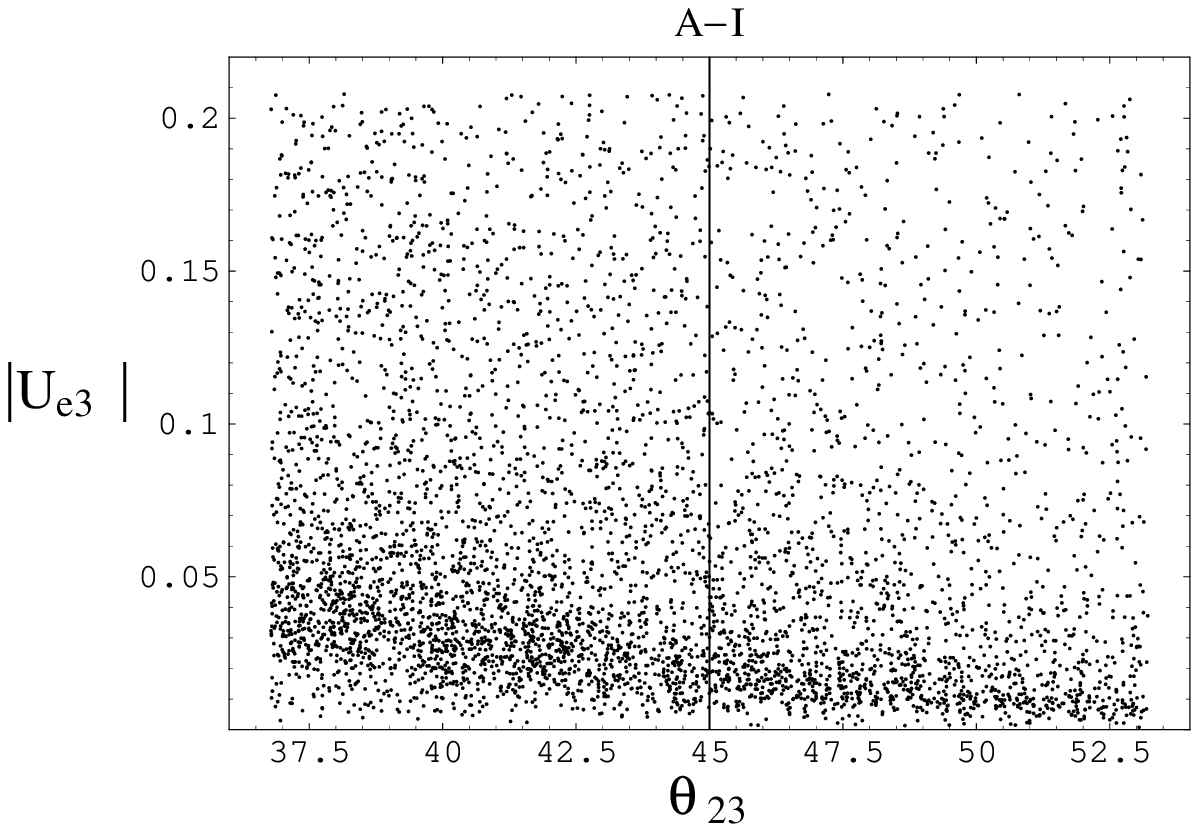}
\hskip 0.5 cm
\epsfxsize=7.0 cm
\epsfbox{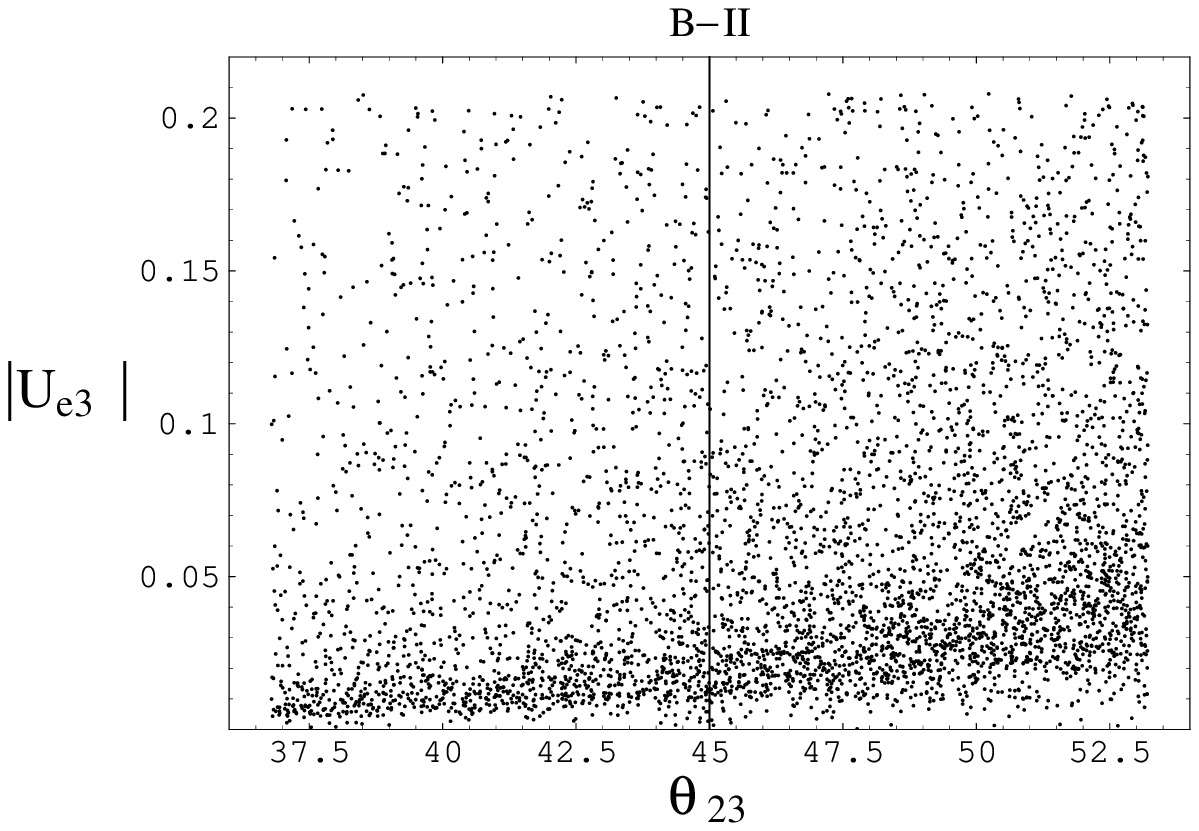}
\end{center}
\vspace*{.0cm}
\caption{The interchange symmetry between A-I and
 B-II, which  corresponds  
$\theta_{23} \leftrightarrow \frac{\pi}{2}-\theta_{23}$.
The horizontal axis is $\theta_{23}$[degrees].}
\end{figure}%

\section{Summary}
 
We presented  numerical and comprehensive analyses of the hybrid textures
of neutrinos, which  may be derived  from the discrete symmetry.
Only six textures among sixty ones  are excluded by the present experimental
data. Since there are many textures  which give similar predictions,
we classified the texture based on  the numerical results.
The neutrinoless double beta decay is also studied in these textures.
The results may disappoint us because we cannot expect
the experimental data to determine the correct texture.
However, it should be  emphasized that there still remain 
 rich structure of the neutrino mass matrix in the phenomenological
 point of view.
We expect that the cooperation between theoretical study 
such as the flavor symmetry and the future neutrino experiments
will reveal the structure of the neutrino mass matrix.

\newpage

 We are most grateful to E. Ma  and M. Frigerio for appropriate advice and
 encouragement.
We also thanks K. Yoshioka for useful  discussions.
  M. T is  supported by the Grant-in-Aid for Science Research,
 Ministry of Education, Science and Culture, Japan(No.16028205). 
S. K is also supported by JSPS.

\bibliographystyle{unsrt}

\newpage
\begin{table}[ht]
 \begin{center}
 \setlength{\tabcolsep}{1.8pt}
  \begin{tabular}{|c|c|c|c|c|c|c|}
   \hline 
    & I & II & III & IV & V & VI \\
   \hline 
      A
    & $\symmatrix{X}{0}{e}{X}{f}{c}$
    & $\symmatrix{X}{d}{0}{X}{f}{c}$
    & $\symmatrix{X}{d}{e}{X}{0}{c}$
    & two zeros
    & two zeros
    & $\symmatrix{X}{d}{e}{X}{f}{0}$ \\
   \hline
      B
    & $\symmatrix{X}{0}{e}{b}{f}{X}$
    & $\symmatrix{X}{d}{0}{b}{f}{X}$
    & $\symmatrix{X}{d}{e}{b}{0}{X}$
    & two zeros
    & $\symmatrix{X}{d}{e}{0}{f}{X}$
    & two zeros \\ 
   \hline
      C
    & $\symmatrix{a}{0}{e}{X}{f}{X}$
    & $\symmatrix{a}{d}{0}{X}{f}{X}$
    & $\symmatrix{a}{d}{e}{X}{0}{X}$
    & $\symmatrix{0}{d}{e}{X}{f}{X}$
    & two zeros
    & two zeros \\ 
   \hline
      D
    & two zeros
    & two zeros
    & $\symmatrix{a}{X}{X}{b}{0}{c}$
    & $\symmatrix{0}{X}{X}{b}{f}{c}$
    & $\symmatrix{a}{X}{X}{0}{f}{c}$
    & $\symmatrix{a}{X}{X}{b}{f}{0}$ \\ 
   \hline
      E
    & two zeros
    & $\symmatrix{a}{X}{0}{b}{X}{c}$
    & two zeros
    & $\symmatrix{0}{X}{e}{b}{X}{c}$
    & $\symmatrix{a}{X}{e}{0}{X}{c}$
    & $\symmatrix{a}{X}{e}{b}{X}{0}$ \\ 
   \hline
      F
    & $\symmatrix{a}{0}{X}{b}{X}{c}$
    & two zeros
    & two zeros
    & $\symmatrix{0}{d}{X}{b}{X}{c}$
    & $\symmatrix{a}{d}{X}{0}{X}{c}$
    & $\symmatrix{a}{d}{X}{b}{X}{0}$ \\ 
   \hline
      G
    & two zeros
    & $\symmatrix{X}{X}{0}{b}{f}{c}$
    & $\symmatrix{X}{X}{e}{b}{0}{c}$
    & two zeros
    & $\symmatrix{X}{X}{e}{0}{f}{c}$
    & $\symmatrix{X}{X}{e}{b}{f}{0}$ \\ 
   \hline
      H
    & $\symmatrix{X}{0}{X}{b}{f}{c}$
    & two zeros
    & $\symmatrix{X}{d}{X}{b}{0}{c}$
    & two zeros
    & $\symmatrix{X}{d}{X}{0}{f}{c}$
    & $\symmatrix{X}{d}{X}{b}{f}{0}$ \\ 
   \hline
      I
    & $\symmatrix{X}{0}{e}{b}{X}{c}$
    & $\symmatrix{X}{d}{0}{b}{X}{c}$
    & two zeros
    & two zeros
    & $\symmatrix{X}{d}{e}{0}{X}{c}$
    & $\symmatrix{X}{d}{e}{b}{X}{0}$ \\ 
   \hline
      J
    & two zeros
    & $\symmatrix{a}{X}{0}{X}{f}{c}$
    & $\symmatrix{a}{X}{e}{X}{0}{c}$
    & $\symmatrix{0}{X}{e}{X}{f}{c}$
    & two zeros
    & $\symmatrix{a}{X}{e}{X}{f}{0}$ \\ 
   \hline
      K
    & $\symmatrix{a}{0}{X}{X}{f}{c}$
    & two zeros
    & $\symmatrix{a}{d}{X}{X}{0}{c}$
    & $\symmatrix{0}{d}{X}{X}{f}{c}$
    & two zeros
    & $\symmatrix{a}{d}{X}{X}{f}{0}$ \\ 
   \hline
      L
    & $\symmatrix{a}{0}{e}{X}{X}{c}$
    & $\symmatrix{a}{d}{0}{X}{X}{c}$
    & two zeros
    & $\symmatrix{0}{d}{e}{X}{X}{c}$
    & two zeros
    & $\symmatrix{a}{d}{e}{X}{X}{0}$ \\ 
   \hline
      M
    & two zeros
    & $\symmatrix{a}{X}{0}{b}{f}{X}$
    & $\symmatrix{a}{X}{e}{b}{0}{X}$
    & $\symmatrix{0}{X}{e}{b}{f}{X}$
    & $\symmatrix{a}{X}{e}{0}{f}{X}$
    & two zeros \\
   \hline
      N
    & $\symmatrix{a}{0}{X}{b}{f}{X}$
    & two zeros
    & $\symmatrix{a}{d}{X}{b}{0}{X}$
    & $\symmatrix{0}{d}{X}{b}{f}{X}$
    & $\symmatrix{a}{d}{X}{0}{f}{X}$
    & two zeros \\
   \hline
      O
    & $\symmatrix{a}{0}{e}{b}{X}{X}$
    & $\symmatrix{a}{d}{0}{b}{X}{X}$
    & two zeros
    & $\symmatrix{0}{d}{e}{b}{X}{X}$
    & $\symmatrix{a}{d}{e}{0}{X}{X}$
    & two zeros \\
   \hline 
  \end{tabular}
  \caption{Sixty hybrid textures. $X$ denotes two equal elements in all  textures.}
 \end{center}
\end{table}

\begin{table}[ht]
 \begin{center}
 \setlength{\doublerulesep}{0.8pt}
 \setlength{\tabcolsep}{1.8pt}
 \renewcommand{\arraystretch}{1.5}
   \begin{tabular}{|c|c|c|c|c|c|c|}
   \hline 
    & I & II & III & IV & V & VI \\
   \hline 
      A
    & NH + DL
    & NH + DL
    & IH + DL
    & -
    & -
    & ALL \\
   \hline
      B
    & NH + DL
    & NH + DL
    & IH + DL
    & -
    & ALL
    & - \\ 
   \hline
      C
    & ALL
    & ALL
    & IH + DL
    & NH
    & -
    & - \\ 
   \hline
      D
    & -
    & -
    & IH + DL
    & NH
    & ALL
    & ALL \\
   \hline
      E
    & -
    & ALL
    & -
    & $\times$
    & IH + DL
    & IH \\ 
   \hline
      F
    & ALL
    & -
    & -
    & $\times$
    & IH
    & IH + DL \\
   \hline
      G
    & -
    & NH
    & IH
    & -
    & ALL
    & IH + DL \\
   \hline
      H
    & NH
    & -
    & IH
    & -
    & IH + DL
    & ALL \\
   \hline
      I
    & NH + DL
    & NH + DL
    & -
    & -
    & IH + DL
    & IH + DL \\
   \hline
      J
    & -
    & ALL
    & IH
    &  $\times$
    & -
    & ALL \\
   \hline
      K
    & ALL
    & -
    & IH
    & $\times$
    & -
    & ALL \\
   \hline
      L
    & ALL
    & ALL
    & -
    & NH
    & -
    & IH \\
   \hline
      M
    & -
    & ALL
    & IH
    & $\times$
    & ALL
    & - \\
   \hline
      N
    & ALL
    & -
    & IH
    &  $\times$
    & ALL
    & - \\
   \hline
      O
    & ALL
    & ALL
    & -
    & NH
    &IH
    & - \\
   \hline 
  \end{tabular}
 \renewcommand{\arraystretch}{1}
  \caption{Mass patterns for each texture. NH, IH and  DL  denote 
 the normal hierarchy, the inverted one, and the degenerate one, respectively.
 ALL denotes NH+IH+DL, and  $\times$ means that the texture is excluded
by the present experimental data.   }
 \end{center}
\end{table}    
\begin{table}
\hskip 2  cm
\begin{tabular}{  | c| c| c| c| c| } \hline
          &   &   &    &   \\
   A-I&   C-III &  C-IV & G-II &  I-V \\
     &   &  &  &   \\  \hline
        &   &   &  &  \\
  $\left (\matrix{ 1&0&\lambda\cr0& 1& 1 \cr
  \lambda & 1 & 1 \cr}\right ) $\  
&  $\left (\matrix{ 1& 1 &1 \cr 1& 1 & 0 \cr
  1 &  0 & 1 \cr}\right ) $\  
&  $\left (\matrix{ 0& \lambda & \lambda\cr \lambda & 1 &1  \cr
  \lambda  & 1 & 1 \cr}\right ) $\ 
 &  $\left (\matrix{ \lambda&\lambda&0\cr \lambda& 1&1 \cr
  0 & 1& 1\cr}\right ) $\ 
& $\left (\matrix{ 1& 1& 1\cr 1& 0& 1\cr
  1 & 1 & \lambda \cr}\right ) $\ \\
   &   &    & & \\  \hline
\end{tabular} 
 \caption{The relative magnitudes of the matrix elements 
 in the unit of $\lambda\simeq 0.2$. }
\end{table} 
\begin{table}
\hskip 0  cm
\begin{tabular}{  |c| c| c| c| } \hline
          &   &   &      \\
 $10\sim 30$ meV & $ 1\sim 3$ meV &  $0.1\sim 0.3$ meV & 0 \\
     &   &  &     \\  \hline
        &   &   &    \\
   Others & C-I, C-II, E-II, F-I, G-II, H-I, J-II, N-I& L-I, L-II, O-I, O-II 
& all-IV\\
   &   &    &  \\  \hline
\end{tabular} 
 \caption{The classification of the  predicted lower bounds of 
$\langle m \rangle_{ee}$.}
\end{table} 
\end{document}